\documentclass[twocolumn,epjc3]{svjour3}          

\RequirePackage[T1]{fontenc}

\smartqed  

\RequirePackage{graphicx}
\RequirePackage{mathptmx}      
\RequirePackage{flushend}
\RequirePackage[numbers,sort&compress]{natbib}
\RequirePackage[colorlinks,citecolor=blue,urlcolor=blue,linkcolor=blue]{hyperref}
\usepackage{subfigure}
\usepackage{amsmath}
\usepackage{amssymb}
\usepackage[dvipsnames]{xcolor}
\usepackage{soul}

\usepackage{subfigure}

\newcommand{\f}{\left(1-\frac{2\,M}{\sqrt{r^2+a^2}}\right)}

\begin{document}\sloppy

\title{Scalar scattering by black holes and wormholes}

\author{Haroldo C. D. Lima Junior\thanksref{e1,addr1}
        \and
        Carolina L. Benone\thanksref{e2,addr2}
        \and
       Lu\'is C. B. Crispino\thanksref{e3,addr1} 
}

\thankstext{e1}{e-mail: haroldolima@ufpa.br}
\thankstext{e2}{e-mail:benone@ufpa.br}
\thankstext{e3}{e-mail: crispino@ufpa.br}

\institute{Programa de P\'os-Gradua\c{c}\~{a}o em F\'{\i}sica, Universidade 
		Federal do Par\'a, 66075-110, Bel\'em, Par\'a, Brazil\label{addr1}
          \and  
          Campus Universit{\'a}rio Salin{\'o}polis, Universidade Federal do Par{\'a}, 68721-000, Salin{\'o}polis, Par{\'a}, Brazil\label{addr2}
}

\date{Received: date / Accepted: date}

\maketitle

\begin{abstract}
We study the scattering of monochromatic planar scalar waves in a geometry that interpolates between the Schwarzschild solution, regular black holes and traversable wormhole spacetimes. We employ the partial waves approach to compute the differential scattering cross section of the regular black hole, as well as of the wormhole solutions. We compare our full numerical results with the classical geodesic scattering and the glory approximation, obtaining excellent agreement in the appropriate regime of validity of such approximations. We obtain that the differential scattering cross section for the regular black hole case is similar to the Schwarzschild result. Notwithstanding, the results for wormholes can be very distinctive from the black hole ones. In particular, we show that the differential scattering cross section for wormholes considerably decreases at large scattering angles for resonant frequencies.

\end{abstract}

\section{Introduction}
The era of precise measurements in strong gravitational field regime has begun. The gravitational wave measurements performed by the LIGO/VIRGO collaboration provided undoubtful evidence for the existence of black holes in nature, detected in the form of a coalescing binary system~\cite{VL1}. Furthermore, the EHT collaboration imaged the shadow of a black hole at the center of the Messier 87 galaxy (M87)~\cite{EHT}. More recent results about the polarization of light emitted next to the center of M87 indicate that the central black hole is surrounded by a strong magnetic field~\cite{EHT2021}.

Black holes arise naturally as solutions of the Einstein field equations. The first black hole solution obtained within Einstein's theory was the Schwarzschild solution, which represents a static and spherically symmetric black hole in vacuum. The generalization to the rotating black hole case was obtained by Roy Patrick Kerr in 1963, and it is nowadays known as the Kerr solution~\cite{Kerr:1963}.  Black holes coupled to linear (Reissner-Nordstrom black hole~\cite{Reissner,Nordstrom}) and non-linear electromagnetic fields (e. g. the Bardeen black hole~\cite{bardeen}) have also been proposed in the literature. In particular, the latter were proposed to circumvent the singularity problem, and became known as regular black hole solutions.

It is possible to consider a wormhole structure associated to the Schwarzschild solution, known as Einstein-Rosen bridge \cite{Einstein-Rosen}. This bridge is represented by a throat that connects two asymptotically flat regions of the maximal analytical extension of the Schwarzschild solution. The Schwarzschild wormhole is not traversable~\cite{Visser-Book}. Traversable wormholes were originally discussed in Ref.~\cite{MorrisThorne}. Although wormholes are seen as exotic solutions, recently it was shown that they can be sourced by reasonable matter~\cite{SalcedoWormholes,Konoplya:2021hsm}. 

Simpson and Visser proposed a geometry that interpolates between the Schwarzschild black hole and wormhole spacetimes, admitting a regular "black bounce" in the middle of the interpolation branch~\cite{Black_bounce}. The wormhole branch of interpolation represents traversable wormholes in the sense discussed in Ref.~\cite{MorrisThorne}. This Simpson-Visser spacetime possesses several interesting properties, which have been recently investigated (cf., e.g., Ref.~\cite{SV-0,SV-1,SV-2,SV-3,SV-4,SV-5,SV-6,SV-7}).

Scattering processes were important to many groundbreaking discoveries in Physics. The discovery of the atomic nucleus itself was achieved through the analysis of the so-called Rutherford scattering, i.e. the scattering of charged particles by the Coulomb force. In General Relativity, scattering processes are also important, for instance, in the analysis of gravitational wave scattering by compact objects. The scattering of waves by static and stationary black holes has been studied in several papers and books~\cite{sct-1,sct-2,sct-3,sct-4,Futterman::book,sct-5,Leite:2019zqo,CDHO2014,MOC2015,CDHO2015,sct-6}. 
However, few investigations dealing with wave scattering by wormhole geometries were presented so far (cf., e.g., Ref.~\cite{dmoc2019}). 

We provide the scattering analysis of scalar waves in wormhole and regular black hole geometries represented by the Simpson-Visser metric, and compare both cases. The analysis of the absorption process for this geometry was presented in Ref.~\cite{SV-1}, where it has been shown that wormholes and black holes can exhibit quite distinctive absorption spectra.
The Simpson-Visser spacetime is represented by the following line element:
\begin{align}
\label{Lineel} ds^2=-f(r)\,dt^2+f(r)^{-1}\,dr^2
+\left(r^2+a^2\right)\,d\Omega^2,
\end{align} 
where 
\begin{align}
&f(r)\equiv\left(1-\frac{2\,M}{\sqrt{r^2+a^2}}\right),
\end{align}
and $d\Omega^2$ is the line element of a unit sphere. The line element \eqref{Lineel} may describe the standard Schwarzschild solution ($a=0$), regular black holes ($0<a<2\,M$) or wormhole solutions ($a\geq 2M$).~(See Ref.~\cite{Black_bounce} for further details.) The black hole event horizon is located at
\begin{equation}
\label{EH}r_h=\sqrt{(2\,M)^2-a^2},
\end{equation}
reducing to the well known Schwarzschild result for $a=0$. 

We study the scattering of planar scalar waves in black holes, as well as wormhole spacetimes, by varying the parameter $a$ present in the Simpson-Visser line element~\eqref{Lineel}.

The remainder of this paper is organized as follows: Section~\ref{section1} is dedicated to the classical scattering cross section, obtained by analyzing the geodesics in the Simpson-Visser spacetime. In Sec.~\ref{section2} we outline the partial waves approach for the scattering of scalar waves. In Sec.~\ref{section3} we present a selection of our numerical results for the black hole and wormhole cases. Our final remarks are presented in Sec.~\ref{final remarks}. We adopt units such that $G=c=\hslash=1$ and metric signature $(-,+,+,+)$.

\section{Classical scattering}
\label{section1}
\subsection{Geodesic scattering}
We study the classical scattering cross section in the Simpson-Visser spacetime, investigating null geodesics propagating in this geometry. Some properties of particle motion in the Simpson-Visser geometry were studied, for instance, in Refs.~\cite{SV-1,SV-2,SV-3,SV-4,SV-5}.  The Lagrangian for null geodesic motion is given by 

\begin{equation}
\label{Lgeo}\mathcal{L}_{geo}=\frac{1}{2}g_{\mu\nu}\dot{x}^\mu\dot{x}^\nu=0,
\end{equation}
where the overdots represent differentiation with respect to the affine parameter along the geodesics. We may restrict our analysis to the equatorial plane $\theta=\pi/2$, since the spacetime is spherically symmetric. Moreover, we note the existence of the following conserved quantities:
\begin{align}
\label{Energy}&E=\f\,\dot{t},\\
\label{Angmomentum}&L=\left(r^2+a^2\right)\,\dot{\varphi},
\end{align}
which are the energy and angular momentum measured by an asymptotic observer, respectively. Using Eqs.~\eqref{Lgeo}-\eqref{Angmomentum}, we may write the equation of motion for massless particles as
\begin{align}
&\label{rdot}\dot{r}^2+V_{\text{eff}}=E^2,\\
&V_{\text{eff}}=\left(1-\frac{2M}{\sqrt{r^2+a^2}}\right)\frac{L^2}{\left(r^2+a^2\right)},
\end{align}
where $V_{\text{eff}}$ is the effective potential for null geodesics. The radius of the closed circular photon orbits ($r_c$) and the corresponding impact parameter ($b_c$) are given by~\cite{SV-1}:
\begin{align}
r_c=\begin{cases}
\sqrt{9M^2-a^2}, \ \ \quad &\text{if \ \ } 0\leq a <2M,\\
\pm \sqrt{9M^2-a^2}, &\text{if \ \ } 2M \leq a \leq 3M,\\
0, &\text{if \ \ } a>3M,
\end{cases}
\end{align}
and
\begin{align}
b_c=\begin{cases}
3\sqrt{3}M, \quad &\text{if \ \ } 0\leq a \leq 3M,\\
\frac{a^{\frac{3}{2}}}{\left(a-2M\right)^\frac{1}{2}}, &\text{if \ \ } a>3M,
\end{cases}
\end{align}
respectively.
Using a chain rule, we may write Eq.~\eqref{rdot} with derivatives with respect to the $\varphi$ coordinate. Moreover, we define the variable 
\begin{equation}
u \equiv \frac{1}{\sqrt{r^2+a^2}},
\end{equation}
and we find that the equation for the orbit of massless particles in the Simpson-Visser spacetime can be written as:
\begin{equation}
\label{orbit_eq}\left(\frac{du}{d\varphi}\right)^2=\left(1-a^2u^2\right)\left[\frac{1}{b^2}-u^2\left(1-2Mu\right)\right]\equiv U(u),
\end{equation}
where $b=L/E$ denotes the impact parameter of the photon. Eq.~\eqref{orbit_eq} reduces to the well known Schwarzschild result for $a=0$. 

The scattering angle is given by
\begin{equation}
\label{scat-angle}\chi(b)=2\,\int^{u_0}_0\frac{du}{\sqrt{U(u)}}-\pi,
\end{equation}
where $u_0$ is related to the turning point of the photon with impact parameter $b$. The turning point is obtained by solving:
\begin{align}
U(u_0)=0.
\end{align} 
The classical scattering cross section is given by \cite{Collins:1973xf}

\begin{eqnarray}
\label{classical-scattering-formula}\frac{d\sigma_{cl}}{d\Omega}=\frac{1}{\sin\chi}\sum b(\chi)\left|\frac{db(\chi)}{d\chi}\right|,
\end{eqnarray}
where the impact parameter $b(\chi)$ in terms of the scattering angle can be obtained by inverting Eq.~\eqref{scat-angle}. The sum in Eq. (\ref{classical-scattering-formula}) stands for the fact that geodesics passing close to the critical orbit can go around the black hole many times before being scattered to infinity, therefore, the sum takes into account $\chi, 2\pi-\chi, 2\pi+\chi$ and so on. We can obtain an analytical expression for the scattering cross section for small scattering angles $\chi$. Expanding the term inside the integral of Eq.~\eqref{scat-angle} in powers of $M$ and $a$, up to second order, and integrating the result, we find that
\begin{equation}
\label{chi(b)}\chi(b)\approx\frac{4M}{b}+\frac{(15\pi-16)M^2}{4b^2}+\frac{a^2\pi}{4b^2},
\end{equation}
which agrees with the expression found in Ref.~\cite{SV-2} up to first order in $M$ and second order in $a$. Solving Eq.~\eqref{chi(b)} for $b(\chi)$ and substituting into Eq.~\eqref{scat-angle}, we obtain
\begin{align}
\label{clas-scat-exp}\frac{d\sigma_{cl}}{d\Omega}\approx\frac{16\,M^2}{\chi^4}+\frac{(15\pi-16)M^2}{4\chi^3}+\frac{a^2\pi}{4\chi^3},
\end{align}
for small values of $\chi$. The expression \eqref{clas-scat-exp} is valid for the black hole and wormhole cases. We note that the contribution from the Simpson-Visser parameter appears only in quadratic order on the classical scattering cross section. Thus we expect a similar scattering cross section to the Schwarzschild case for small values of the parameter $a$.

\begin{figure*}
  \centering
  \subfigure{\includegraphics[scale=0.6]{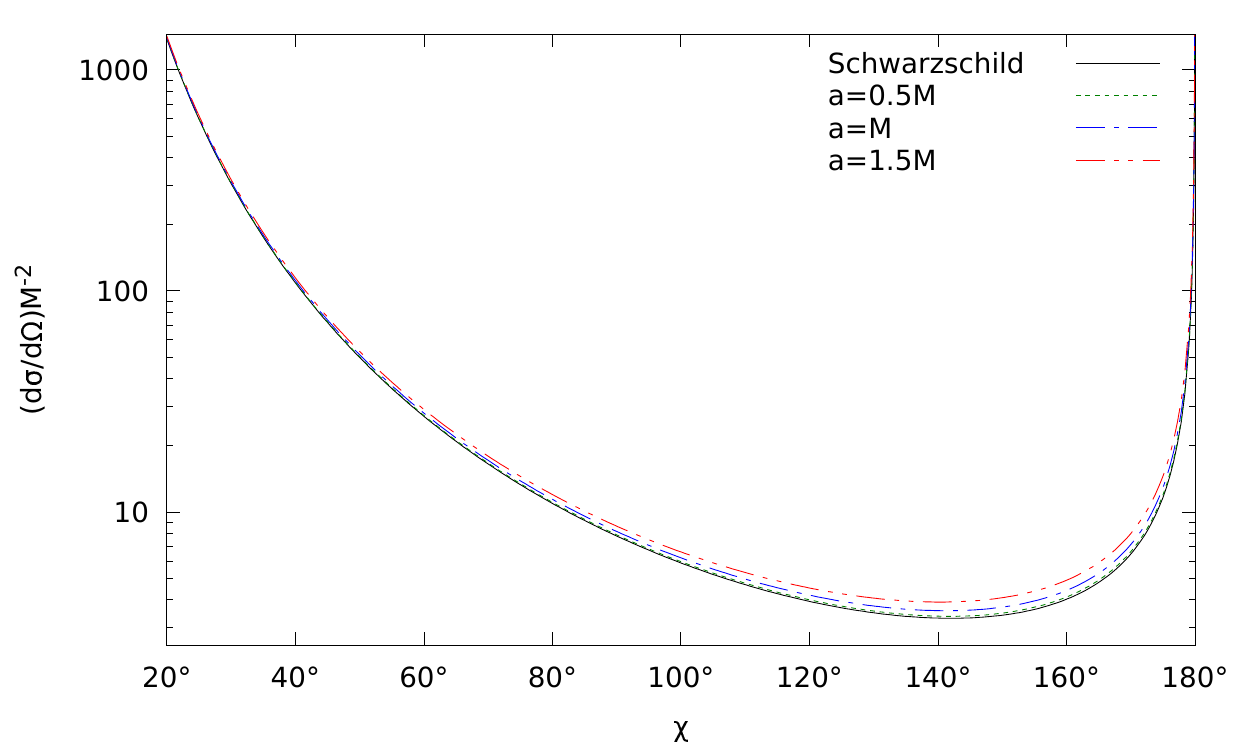}}
  \subfigure{\includegraphics[scale=0.6]{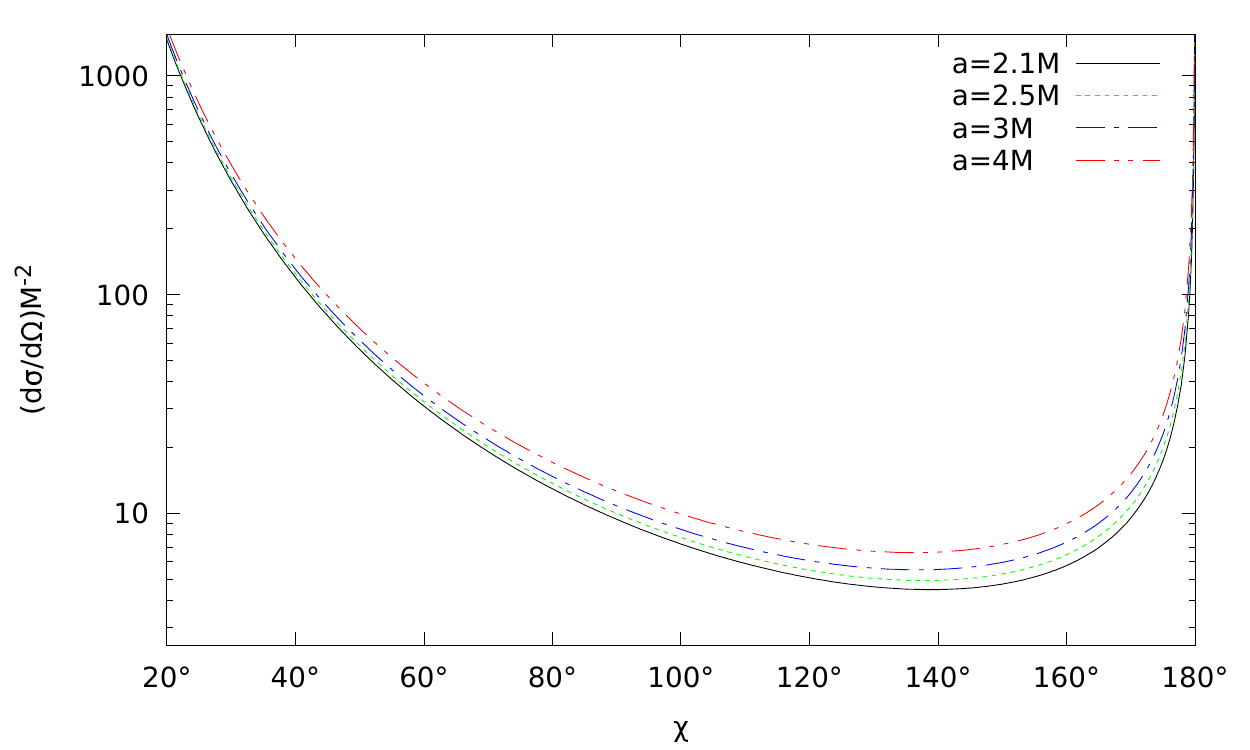}}
  \caption{The classical scattering cross section of massless particles by black holes (left panel) and wormholes (right panel) for different values of $a/M$. For the black hole case, we also show the Schwarzschild result.}
\label{clas-sct}
\end{figure*}

In Fig.~\ref{clas-sct}, we show the classical scattering cross section of the Simpson-Visser geometry, in the black hole as well as in the wormhole ranges of interpolation. We obtained the results shown in Fig.~\ref{clas-sct} numerically: We integrate numerically Eq.~\eqref{scat-angle} for several values of $b$, and compute an interpolation function $\chi(b)$. Then we compute the inverse function $b(\chi)$ and plug into Eq.~\eqref{classical-scattering-formula}. From Fig.~\ref{clas-sct}, we note that the more we increase the value of $a/M$, more the classical scattering cross section 
differs from the Schwarzschild black hole result. More precisely, the classical scattering cross section increases as we increase the value of $a/M$. As shown in Fig.~\ref{clas-sct}, the difference in the classical scattering cross section is more evident for angles around $\chi=150^\circ$. 

\subsection{Glory scattering}
\label{Glory-Section}
\begin{figure}
\subfigure{\includegraphics[scale=0.6]{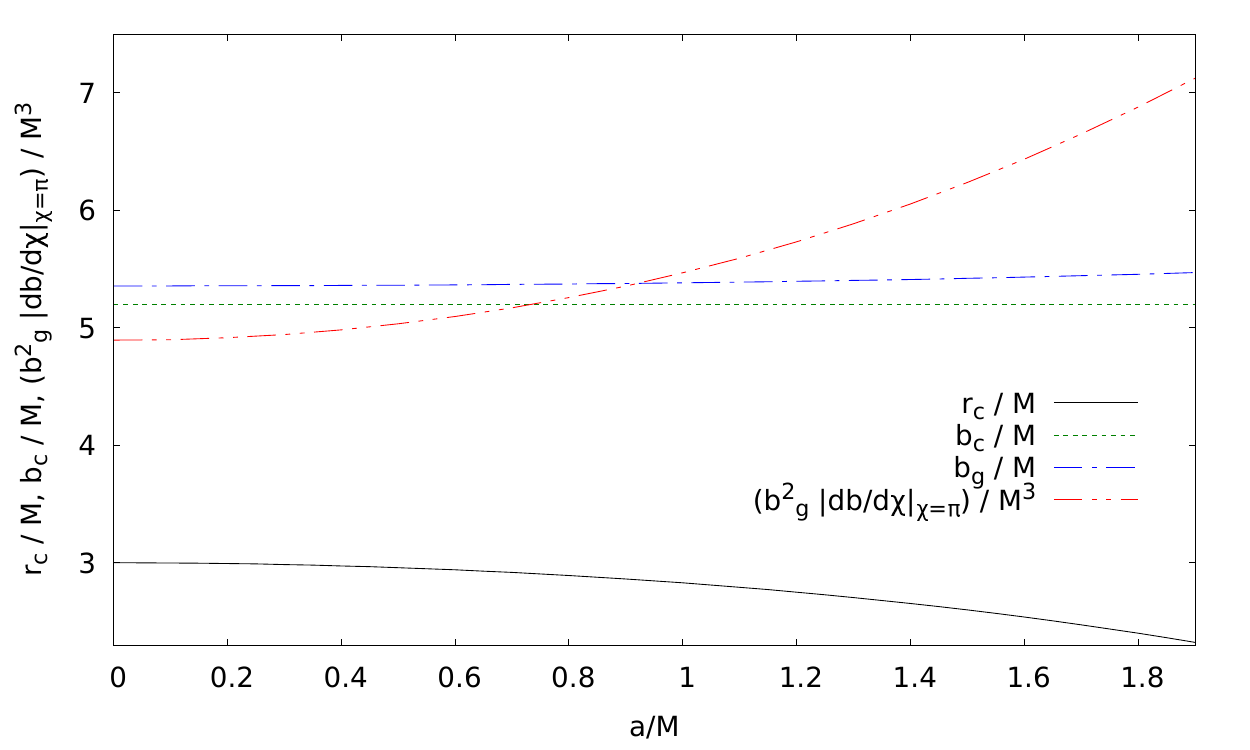}}
\caption{The glory parameters for the Simpson-Visser black hole in terms of the interpolation parameter $a/M$ present in the line element~(\ref{Lineel}). We note that $b_g$ and $b_g^2|db/d\chi|_{\chi=\pi}$ are monotonically increasing functions of the interpolation parameter $a$, while $r_c$ decreases monotonically with $a$. The critical impact parameter $b_c$ remains constant for the whole range of the interpolation parameter of the black hole case.}
\label{glory-terms}
\end{figure}

The classical cross section captures interesting features concerning the motion of massless particles in the Simpson-Visser geometry. However, wave effects, for instance interference, can not be obtained within the classical regime. In order to study semiclassical effects in the Simpson-Visser spacetime, we can apply the glory approximation to the black hole branch of interpolation~\cite{Matzner::1985}. The Simpson-Visser geometry presents an unstable photon orbit, therefore massless particles can be scattered in arbitrarily large angles. The glory approximation is suitable for waves scattered at angles $\chi \approx \pi$. Moreover, since it is a semiclassical approximation, we use it to analyze the scattering of scalar waves in the high-frequency regime ($\omega\,M \gg 1$). The scattering of scalar waves in the low-to-mid frequency regime is presented in Sec.~\ref{section2} within the full numerical approach. The semiclassical glory approximation for black holes is given by~\cite{Matzner::1985}:
\begin{align}
\label{glory-formula}\frac{d\sigma}{d\Omega}=2\pi\omega b_g^2 \left|\frac{d b}{d\chi} \right|_{\chi=\pi} J^2_{2s}(\omega b_g \sin\chi),
\end{align}
where $\omega$ is the frequency of the wave, $b_g$ is the impact parameter for backscattered waves ($\chi \approx \pi$), $J_{2s}$ is the Bessel function of the first kind and $s$ is the spin of the wave. Here we set $s=0$, since we are dealing with scalar waves. In Fig.~\ref{glory-terms}, we plot the main quantities present in the glory formula, as a function of the interpolation parameter $a$ for the black hole case. We note that the critical impact parameter $b_c$ remains constant and equal to the value of the impact parameter of the Schwarzschild black hole ($b_c=3\sqrt{3}M$). This is related to the shadow degeneracy results for the Simpson-Visser spacetime, studied in Ref.~\cite{SV-5}.
The monotonically increasing behavior of $b_g^2|db/d\chi|$ in Fig.~\ref{glory-terms} reveals that the intensity of the backscattered flux is enhanced as we increase the value of $a$. The impact parameter for backscattered waves ($b_g$) slightly increases as we increase the interpolation parameter $a$. Hence we expect that the interference fringes become slightly narrower for higher values of $a$.
 In Sec.~\ref{section3}, we compare the results obtained with the glory approximation and the full numerical method for the black hole case, obtaining excellent agreement for $\chi \approx \pi$.

\section{The partial waves approach}
\label{section2}
In this section we study the scattering of massless scalar waves in the Simpson-Visser spacetime. The dynamics of the massless scalar waves is described by the Klein-Gordon equation:
\begin{equation}
\label{field_eq}\nabla_\mu\nabla^\mu\Phi=\frac{1}{\sqrt{\left|g\right|}}\partial_\mu\left(\sqrt{\left|g\right|}\,g^{\mu\nu}\partial_\nu\Phi\right)=0.
\end{equation}
Since the Simpson-Visser spacetime is spherically symmetric and we are considering monochromatic scalar waves, 
we can decompose the field $\Phi$ as 
\begin{equation}
\label{modes}\Phi=\sum_{l,m}\frac{\phi(r)}{\left(r^2+a^2\right)^\frac{1}{2}}\,Y_{ l m}(\theta,\varphi)\,e^{-i\omega t},
\end{equation}
where $Y_{ l m}$ are the well known scalar spherical harmonics. By substituting Eq.~\eqref{modes} in Eq.~\eqref{field_eq}, we find that $\phi(r)$ obeys a Schr{\"o}dinger-like equation~\cite{SV-1}:
\begin{equation}
\label{rad_eq}\left(\frac{d^2}{dx^2}+\omega^2-V_{\textit{eff}}\right)\,\phi(x)=0,
\end{equation}
where
\begin{align}
\label{veff} &V_{\textit{eff}}\equiv f(r)\,\left[\frac{f'(r)\,r}{\left(r^2+a^2\right)}+\frac{a^2\,f(r)}{\left(r^2+a^2\right)^2}+\frac{l\,(l+1)}{\left(r^2+a^2\right)}\right],
\end{align}
and $x$ is the tortoise coordinate:
\begin{equation}
dx\equiv f^{-1}(r)\,dr.
\end{equation}
The prime in Eq.~\eqref{veff} denotes differentiation with respect to the radial coordinate $r$. We must impose boundary conditions for Eq.~\eqref{rad_eq}. For the scattering/absorption problem such boundary conditions can be defined in terms of the so-called \textit{in modes}~ \cite{BC2016}. For the black hole case, the \textit{in modes} are described by
\begin{equation}
\label{BC_BH}\phi^{bh}(r) \approx \begin{cases} \mathcal{S}_I+R_{\omega l}\mathcal{S}^*_{I}, & r \rightarrow +\infty \ \ (x\rightarrow +\infty),\\
T_{\omega l}\,\mathcal{S}_{II}, & r \rightarrow r_h\ \ \ \ (x \rightarrow -\infty),
\end{cases}
\end{equation}
while for the wormhole case they are given by
\begin{equation}
\label{BC_WMHL}\phi^{w}(r) \approx \begin{cases} \mathcal{S}_I+\mathcal{S}^*_{I}\,R_{\omega l}, & r \rightarrow +\infty \ \ (x\rightarrow +\infty),\\
T_{\omega l}\,\mathcal{S}_I, & r \rightarrow -\infty \ \ \ (x \rightarrow -\infty),
\end{cases}
\end{equation}
where
\begin{align}
&\mathcal{S}_I=e^{-i\omega x}\sum_{i=0}^N \frac{A_{\infty}^i}{r^i},\\
&\mathcal{S}_{II}=e^{-i\omega x}\sum_{i=0}^N(r-r_h)^i\,A_{rh}^i.
\end{align}
In Eqs.~\eqref{BC_BH} and \eqref{BC_WMHL}, $\left|R_{\omega l}\right|^2$ and $\left|\mathcal{T}_{\omega l}\right|^2$ are the scalar wave reflection and transmission coefficients, respectively. Since the flux is conserved, we have $\left|R_{\omega l}\right|^2+\left|\mathcal{T}_{\omega l}\right|^2=1$.

To investigate the scattering of scalar waves in the Simpson-Visser spacetime, we use the standard time-independent scattering theory in the context of General Relativity. The scattering amplitude is given by~\cite{Futterman::book}
\begin{equation}
\label{SA}f(\theta)=\frac{1}{2i\omega}\sum_{l=0}^{\infty}(2l+1)\left[e^{2i\delta_l(\omega)}-1\right]P_l(\cos\theta),
\end{equation}
where $P_l(\cos\theta)$ are the Legendre polynomials, and $\delta_l(\omega)$ are the phase shifts, which are related to $R_{\omega l}$ according to 
\begin{equation}
\label{phase-shift}e^{2i\delta_l(\omega)}\equiv (-1)^{l+1}R_{\omega l}.
\end{equation}
The differential scattering cross section is the squared modulus of the scattering amplitude, namely~\cite{Futterman::book}
\begin{equation}
\label{DSCS}\frac{d\sigma_{sc}}{d\Omega}=\left|f(\theta)\right|^2.
\end{equation}

\begin{figure}
  \centering
  \subfigure{\includegraphics[scale=0.6]{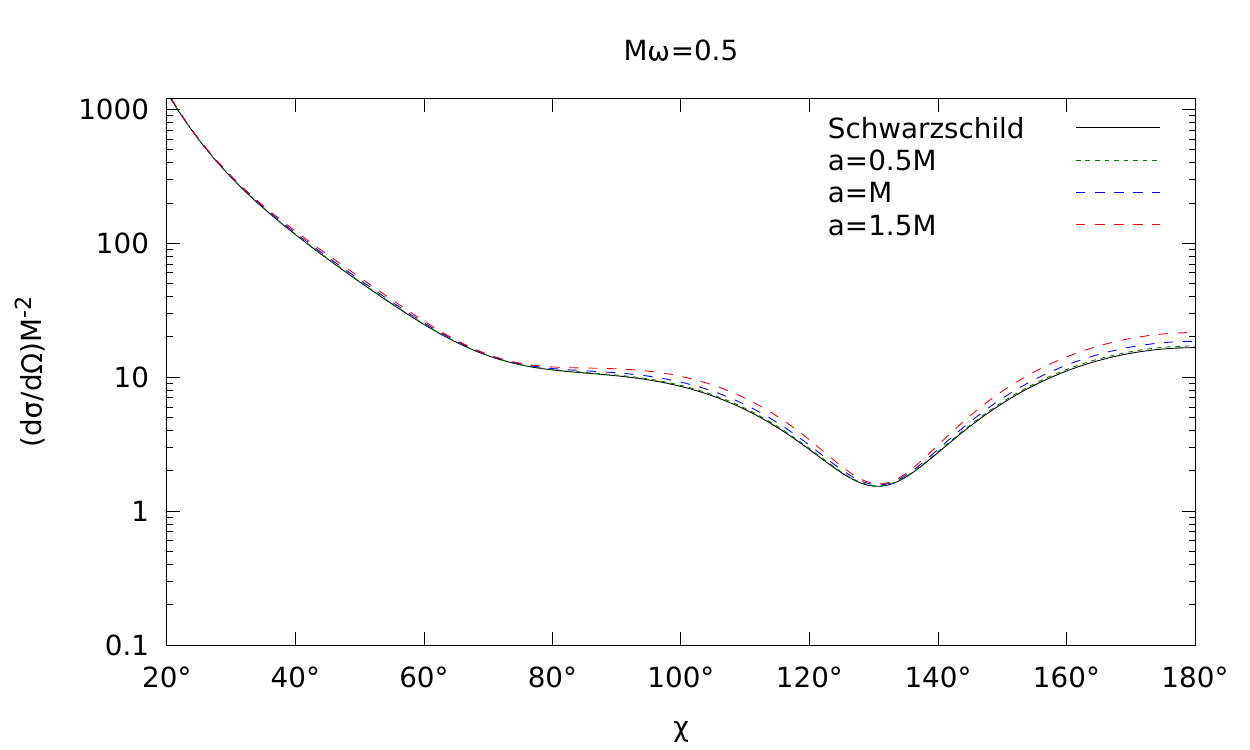}}
  \quad
  \subfigure{\includegraphics[scale=0.6]{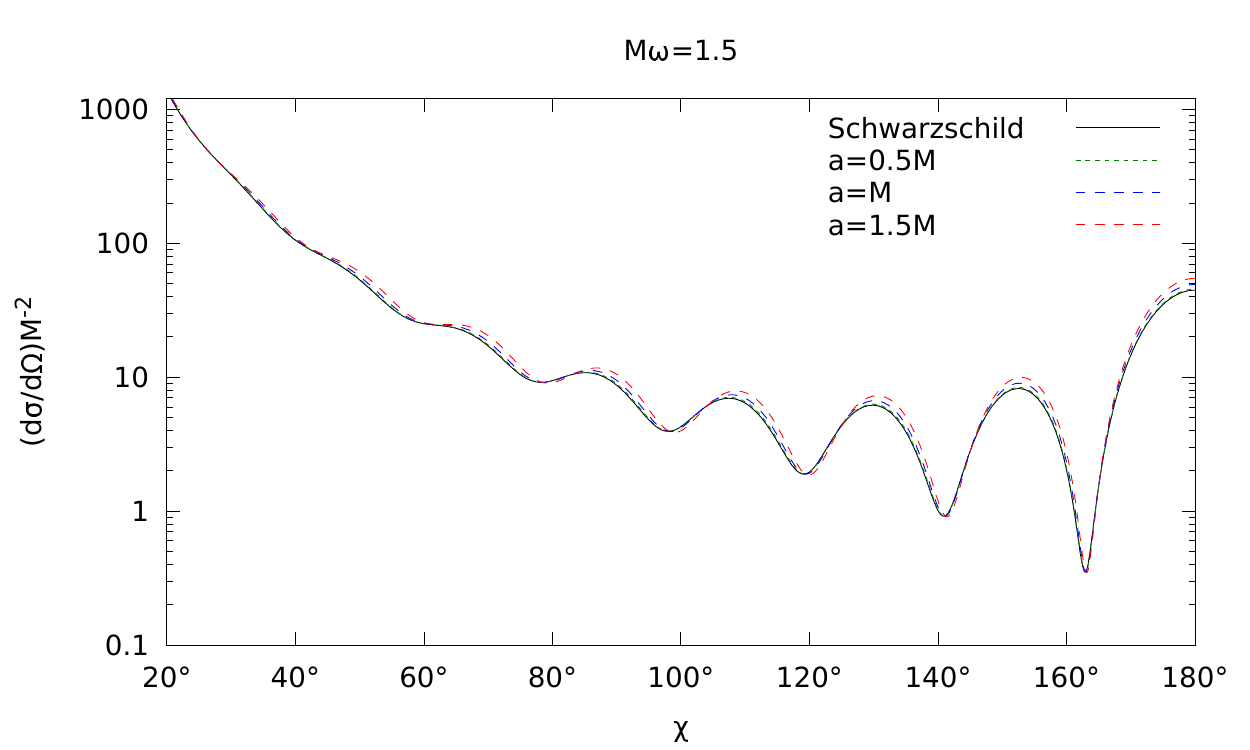}}
\subfigure{\includegraphics[scale=0.6]{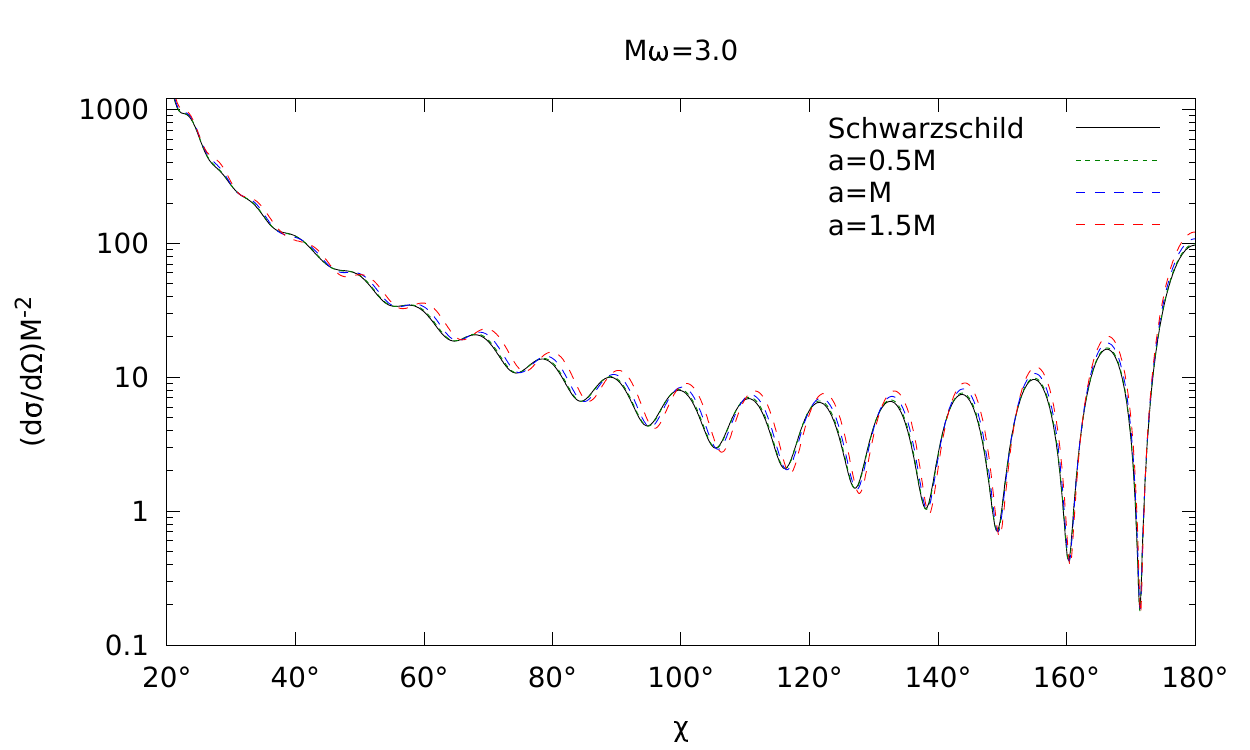}}
\caption{The scalar differential scattering cross section of regular black holes described by the Simpson-Visser metric with different choices of $M\omega$ and $a/M$. We also show the Schwarzschild case for comparison purposes.}
\label{Scattering-BH}
\end{figure}

\begin{figure}
  \centering
  \subfigure{\includegraphics[scale=0.6]{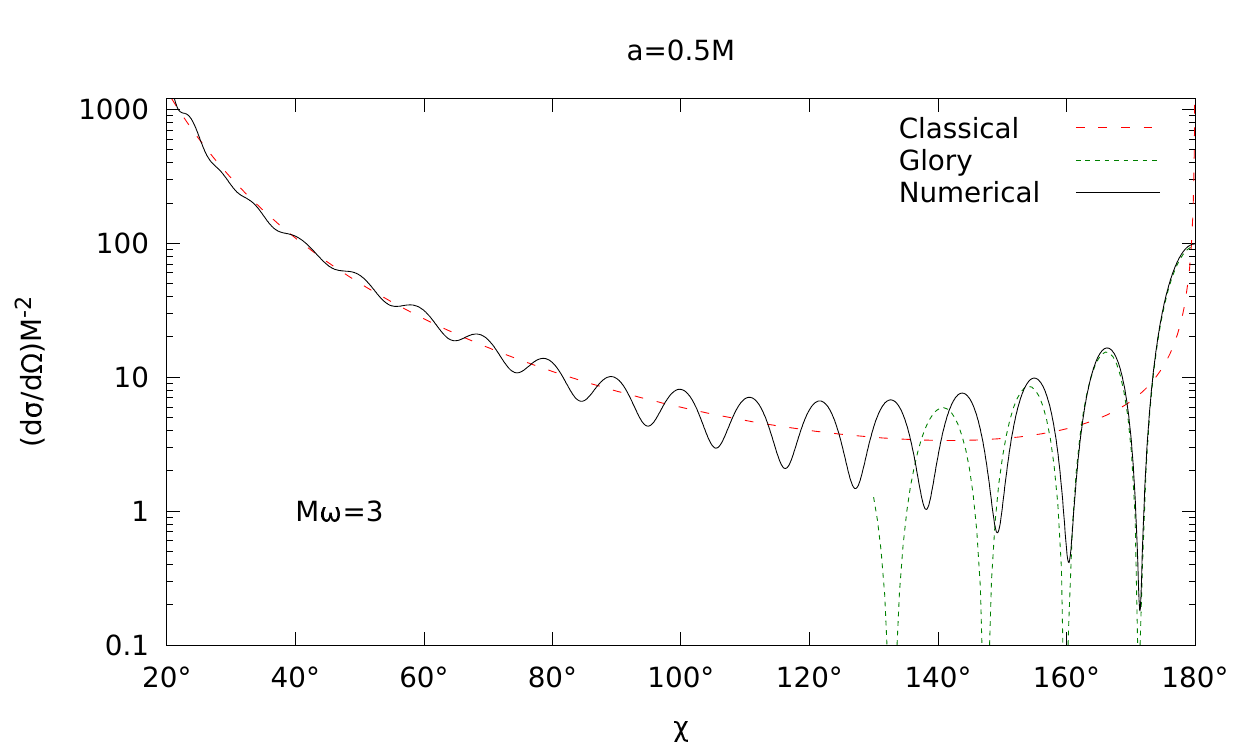}}
  \quad
  \subfigure{\includegraphics[scale=0.6]{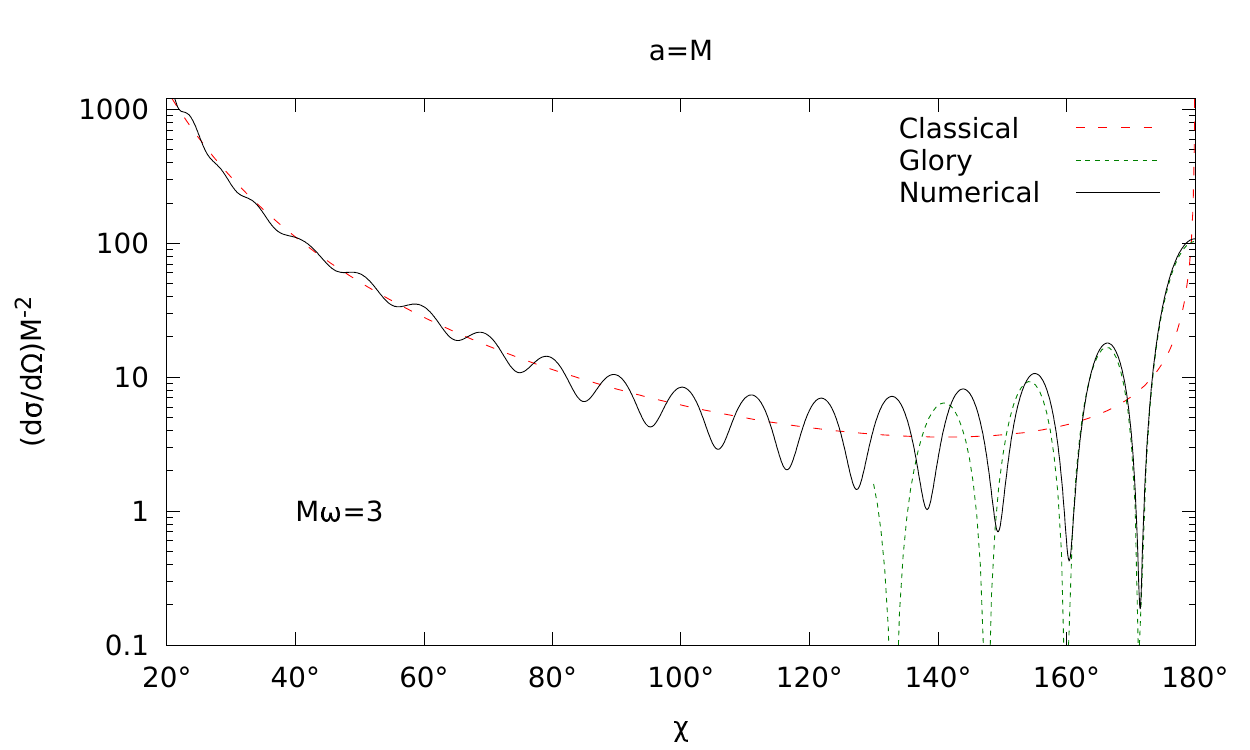}}
\subfigure{\includegraphics[scale=0.6]{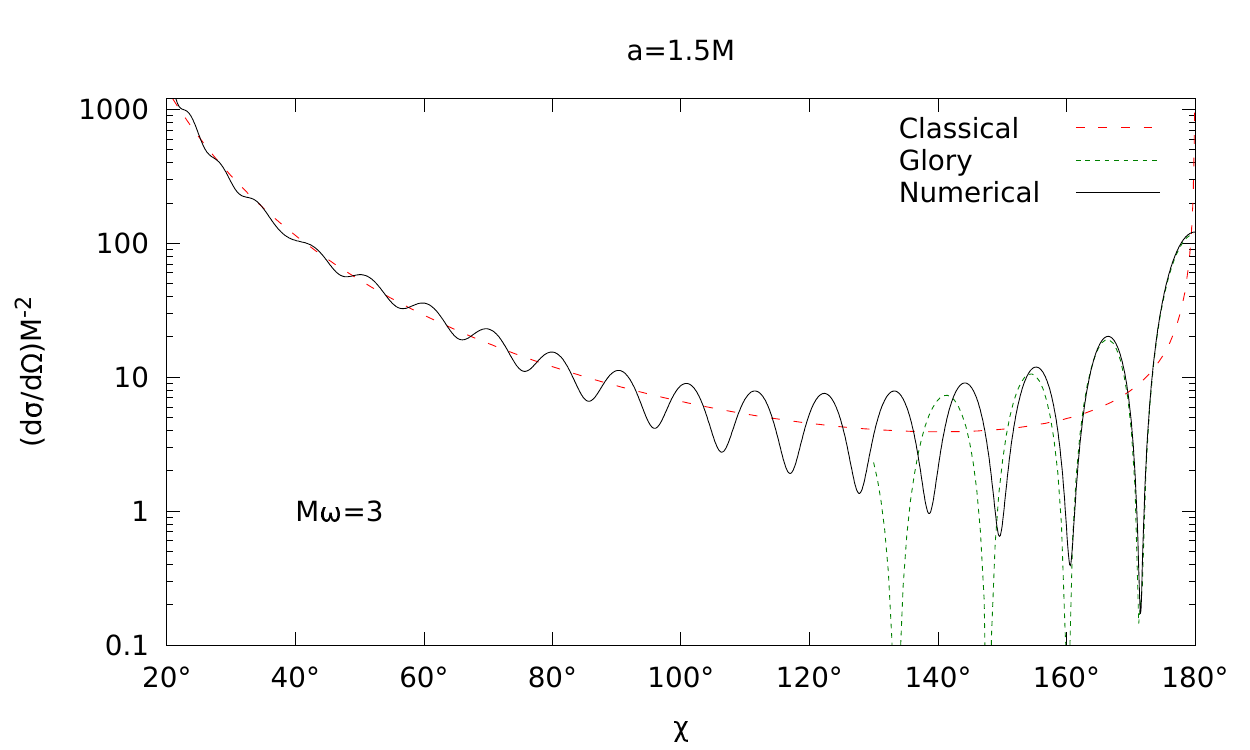}}
\caption{Comparison between the scalar differential scattering cross section of Simpson-Visser regular black holes obtained by the classical scattering, the glory scattering approximation and the full numerical analysis for $M\omega=3$ and different values of $a/M$.}
\label{Scattering-BH2}
\end{figure}

\begin{figure}
  \centering
  \subfigure{\includegraphics[scale=0.6]{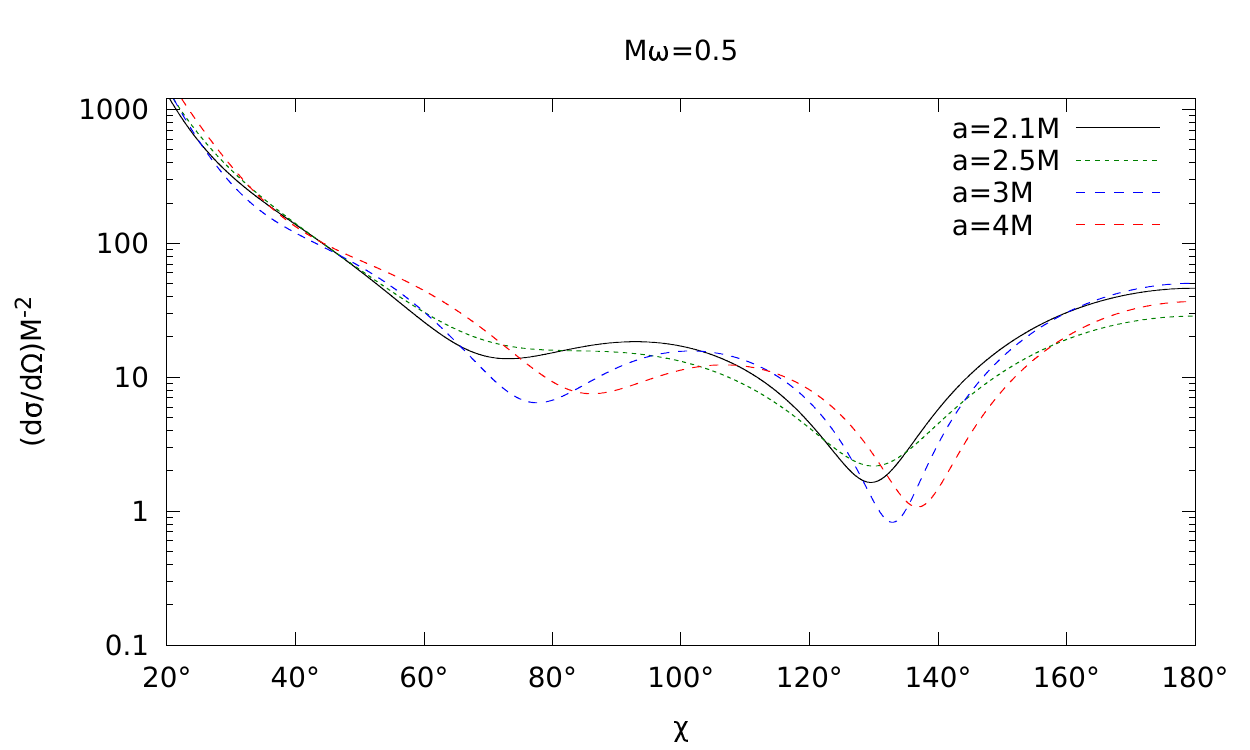}}
  \quad
  \subfigure{\includegraphics[scale=0.6]{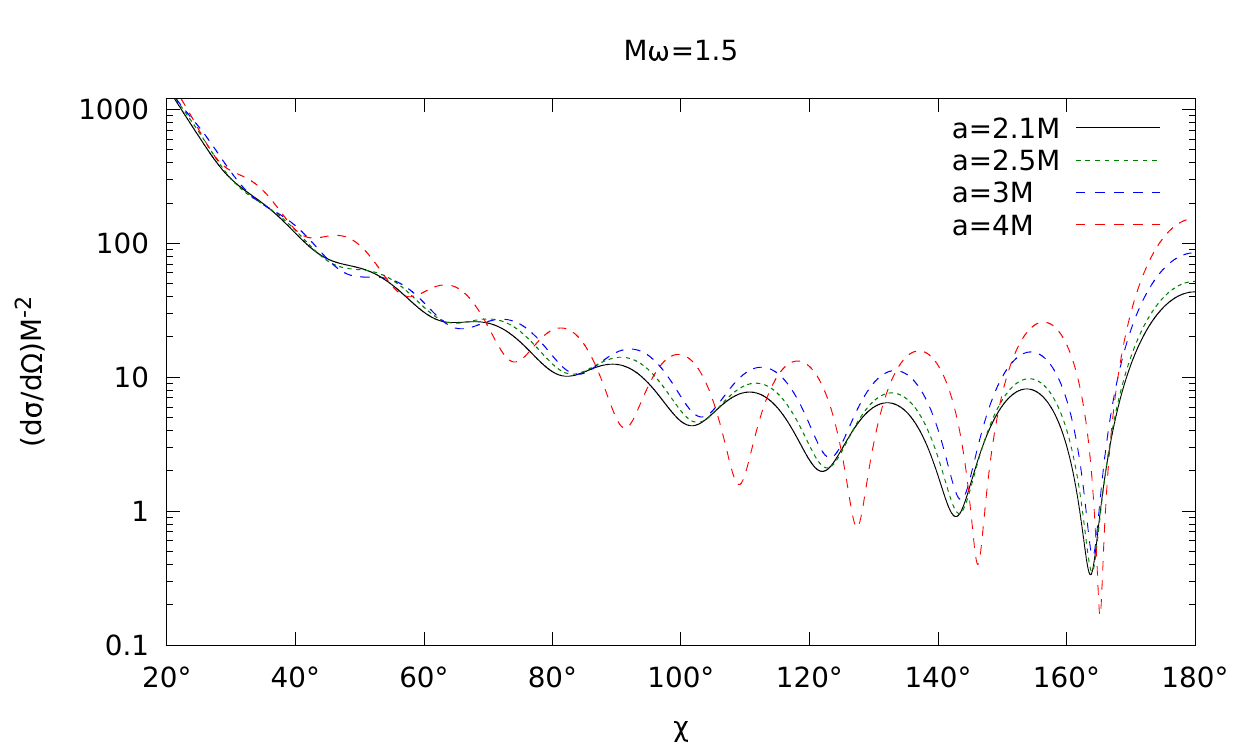}}
\subfigure{\includegraphics[scale=0.6]{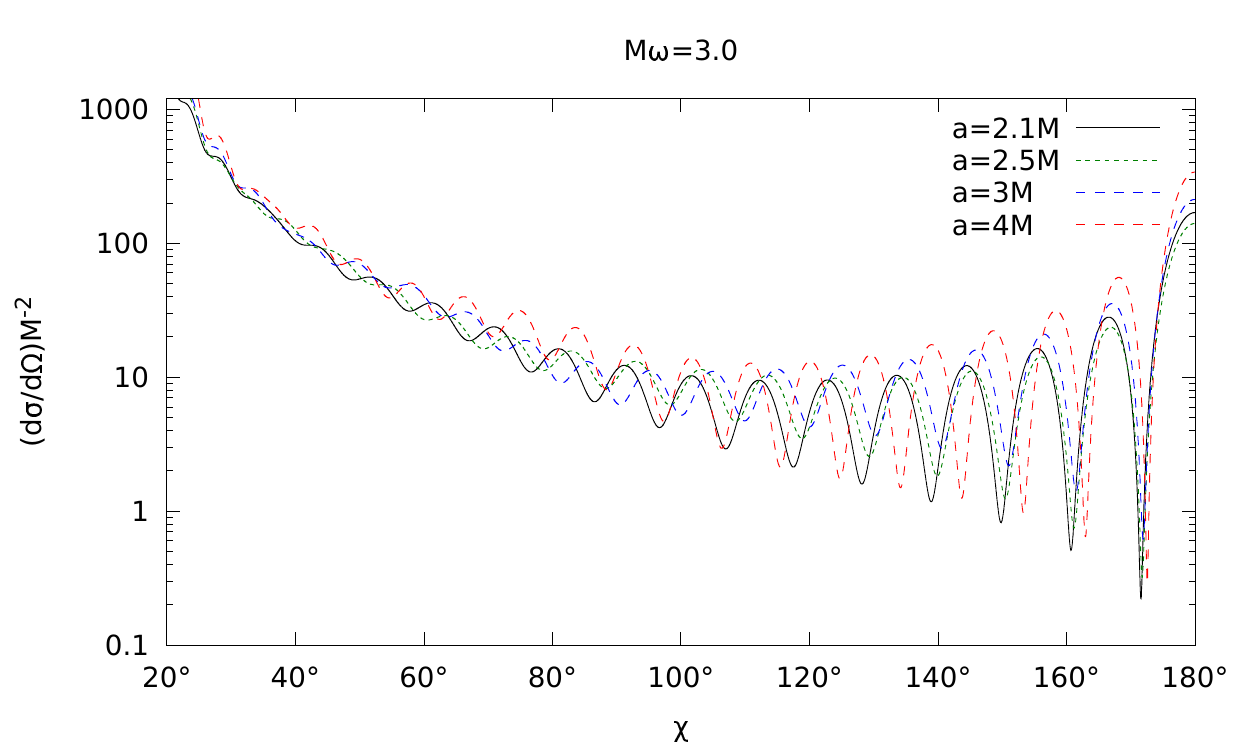}}
\caption{The scalar differential scattering cross section of traversable wormholes described by the Simpson-Visser metric with different values of the frequency and of the interpolation parameter.}
\label{Scattering-WMHL}
\end{figure}

In order to compute the differential scattering cross section through the partial waves method, we solve numerically Eq.~\eqref{rad_eq} subjected to the boundary condition \eqref{BC_BH} in the black hole case, or to the boundary condition \eqref{BC_WMHL} in the wormhole case. From this numerical solution, we compute the reflection coefficient, and hence we find the phase shift and the scattering amplitude from Eqs.~\eqref{phase-shift} and \eqref{SA}, respectively. Due to the poor convergence of the sum in the scattering amplitude $f(\theta)$ for angles $\theta\approx 0$, we need to use the so-called reduced series, outlined for the problem of high-energy scattering of electrons by nuclei~\cite{Yennie::1954}, and first applied in the context of General Relativity in the computation of fermion scattering in the Schwarzschild spacetime~\cite{sct-5}.
Once the scattering amplitude is known, the differential scattering cross section is computed using Eq.~\eqref{DSCS}. A selection of our numerical results is presented in Sec.~\ref{section3} for the black hole, as well as for the wormhole cases.

\section{Numerical results}
\label{section3}
\subsection{Black hole results}
In Fig.~\ref{Scattering-BH} we show the full numerical results for the scattering of planar massless scalar waves by black holes for different values of the interpolation parameter $a$ and $M\omega= 0.5, 1.5$ and $3.0$. We also show the Schwarzschild results for comparison. From Fig.~\ref{Scattering-BH} we note that the change of the parameter $a$ implies in a subtle change on the scattering cross section, implying that it is hard to distinguish the scattering of scalar waves by the regular Simpson-Visser black holes and by the Schwarzschild black hole. A similar conclusion was obtained in the classical scattering regime using backwards ray-tracing in Ref.~\cite{SV-5}.
In Fig.~\ref{Scattering-BH2} we present a comparison between the classical scattering, the glory approximation and the full numerical partial waves method for different values of $a$ and $M\omega=3$. We note that the numerical results oscillate around the classical ones, as a consequence of the interference of waves that orbit the black hole in opposite senses. Moreover, the classical scattering and the full numerical results agree very well in the small scattering angle regime. From Fig.~\ref{Scattering-BH2} we note that the numerical results and the glory approximation are in very well agreement for $\chi\approx \pi$.

\subsection{Wormhole results}
\begin{figure}
  \centering
  \subfigure{\includegraphics[scale=0.6]{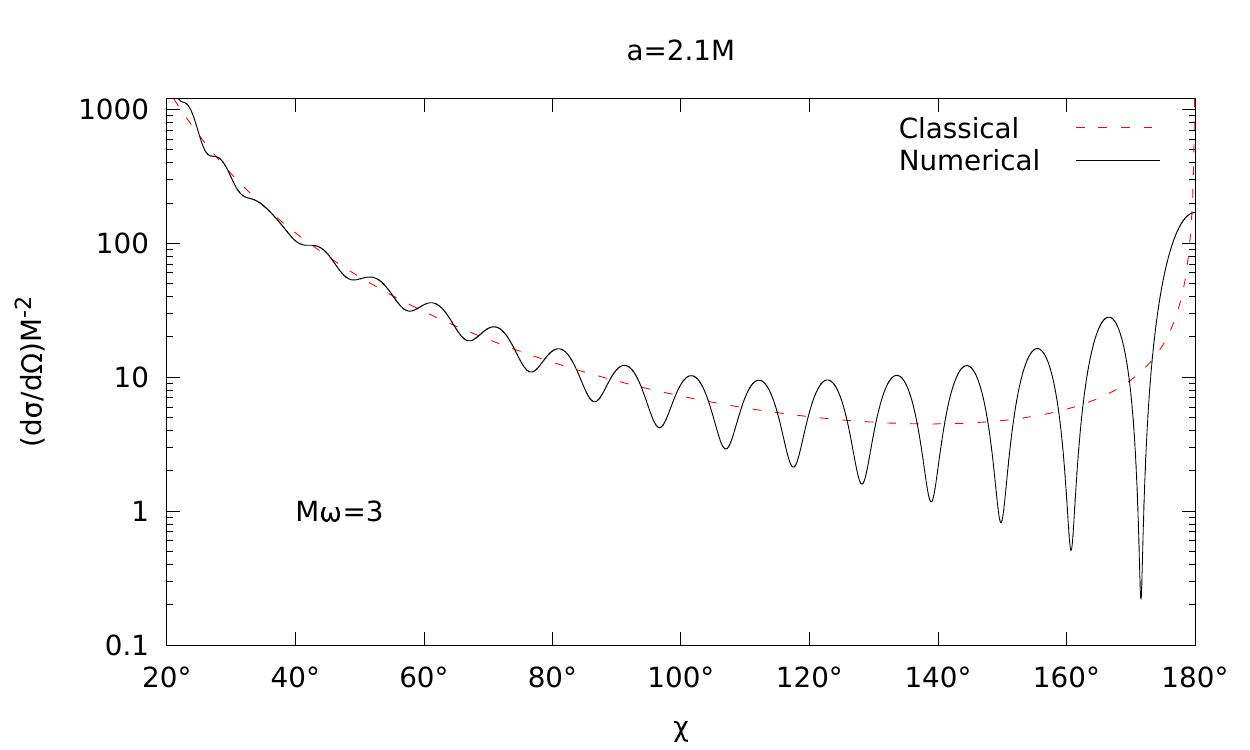}}
  \quad
  \subfigure{\includegraphics[scale=0.6]{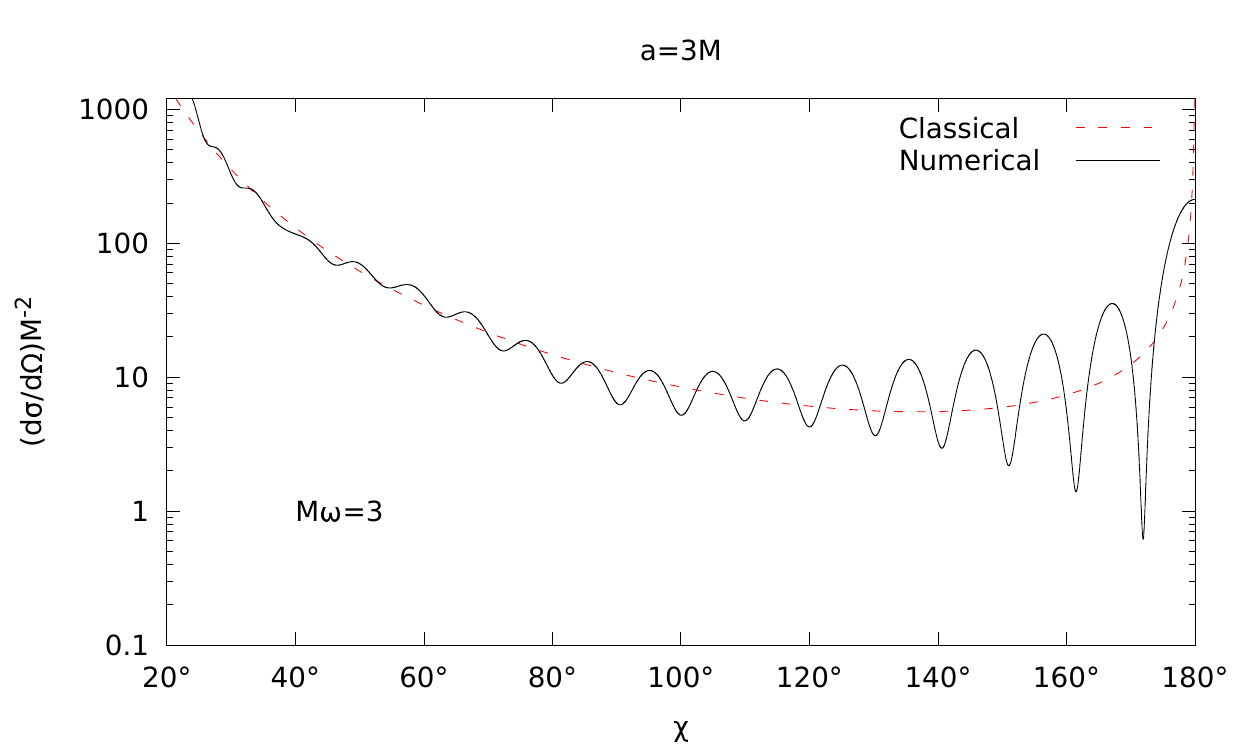}}
\subfigure{\includegraphics[scale=0.6]{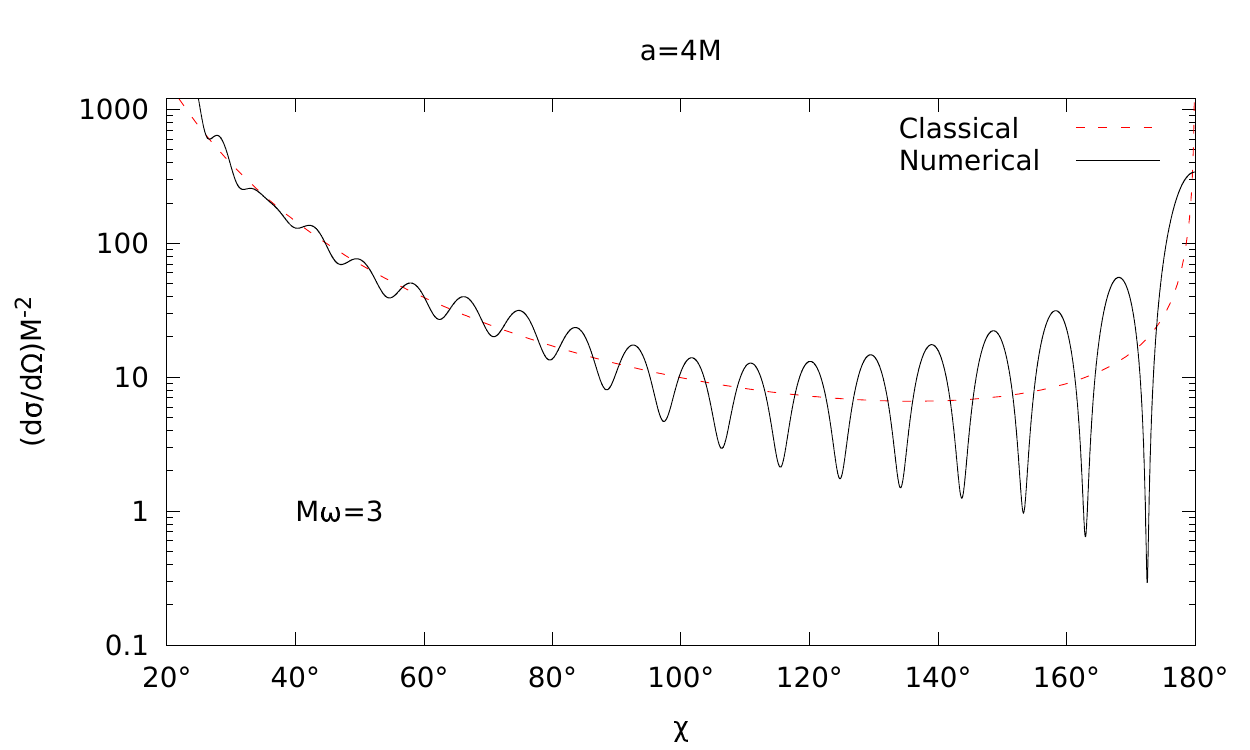}}
\caption{Comparison between the scalar differential scattering cross section of traversable wormholes obtained with the full numerical analysis and the classical result,  for $M\omega=3$ and different values of $a/M$.}
\label{Scattering-WMHL2}
\end{figure}

In Fig.~\ref{Scattering-WMHL} we show the full numerical results for the planar massless scalar waves differential scattering cross section of wormholes with different values of the parameter $a$ and $M\omega= 0.5, 1.5$ and $3.0$. We note that the wormhole case, like the black hole case, present a divergence in the forward direction ($\chi\approx 0^\circ$) and also a glory in the backward direction ($\chi\approx 180^\circ$).
The differential scattering cross section can be quite distinctive for different values of the parameter $a$, in the wormhole branch of interpolation. This contrasts with the black hole case, for which the scattering cross section is very similar for $0\leq a<2M$. In Fig.~\ref{Scattering-WMHL2}, we show the comparison between the classical scattering cross section and the full numerical results. We notice that the numerical results obtained with the partial waves approach oscillates around the classical results, as expected. 

\begin{figure}
  \centering
  \subfigure{\includegraphics[scale=0.6]{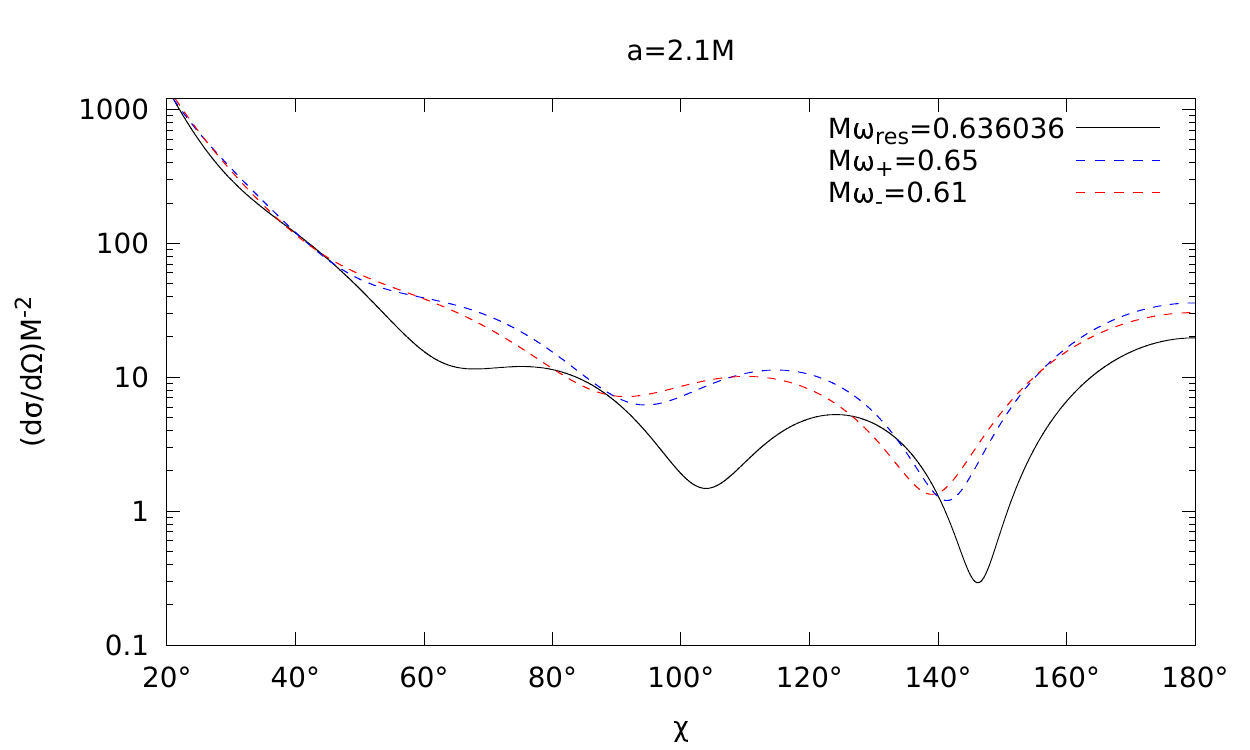}}
  \quad
  \subfigure{\includegraphics[scale=0.6]{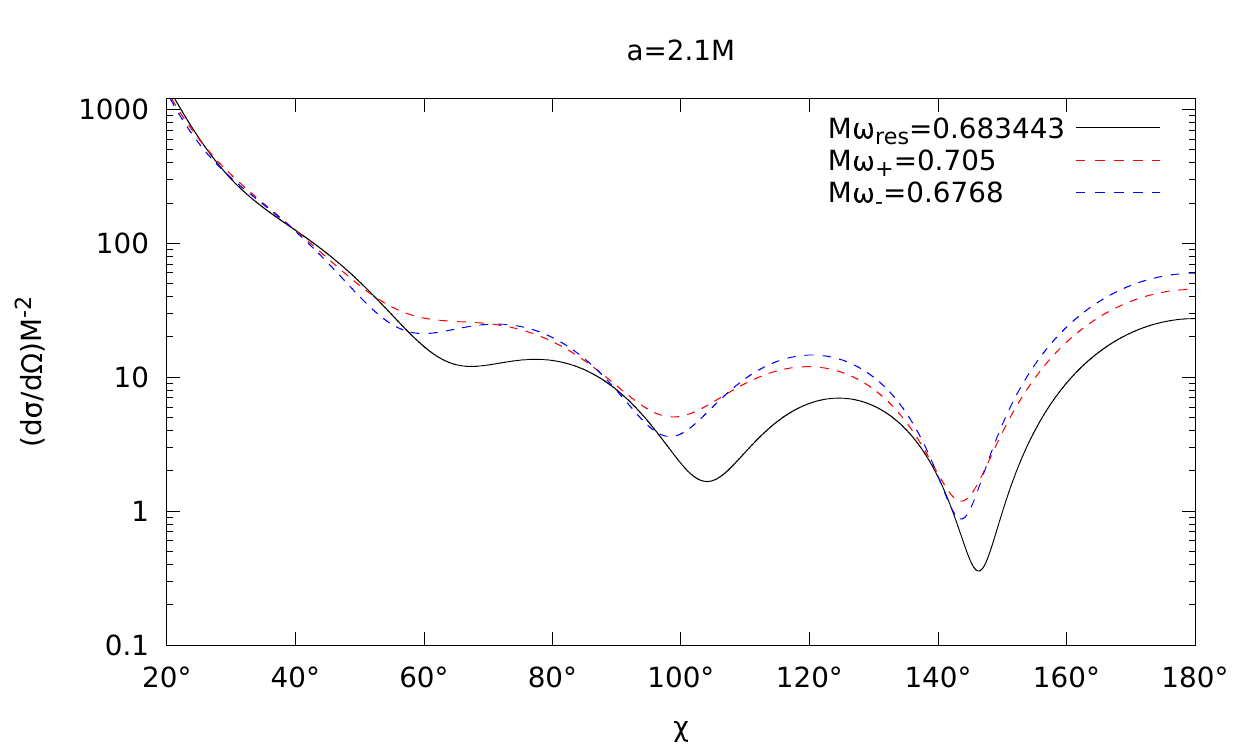}}
\subfigure{\includegraphics[scale=0.6]{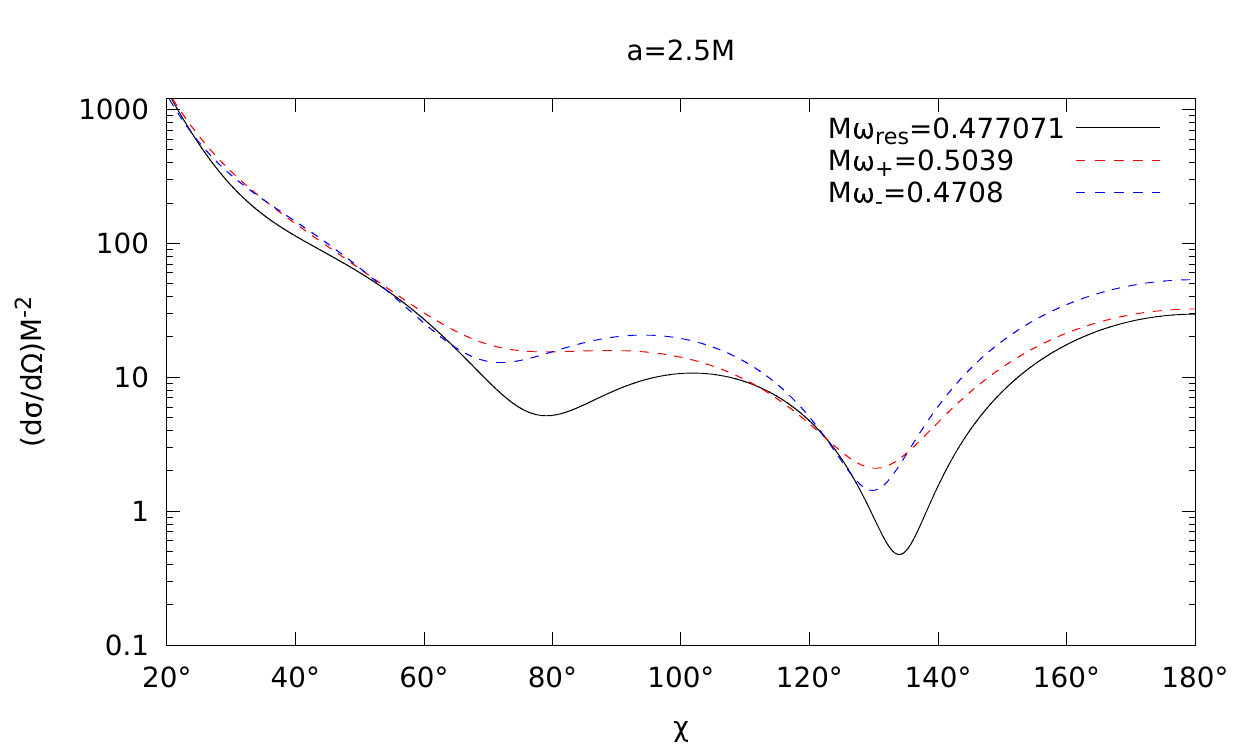}}
\subfigure{\includegraphics[scale=0.6]{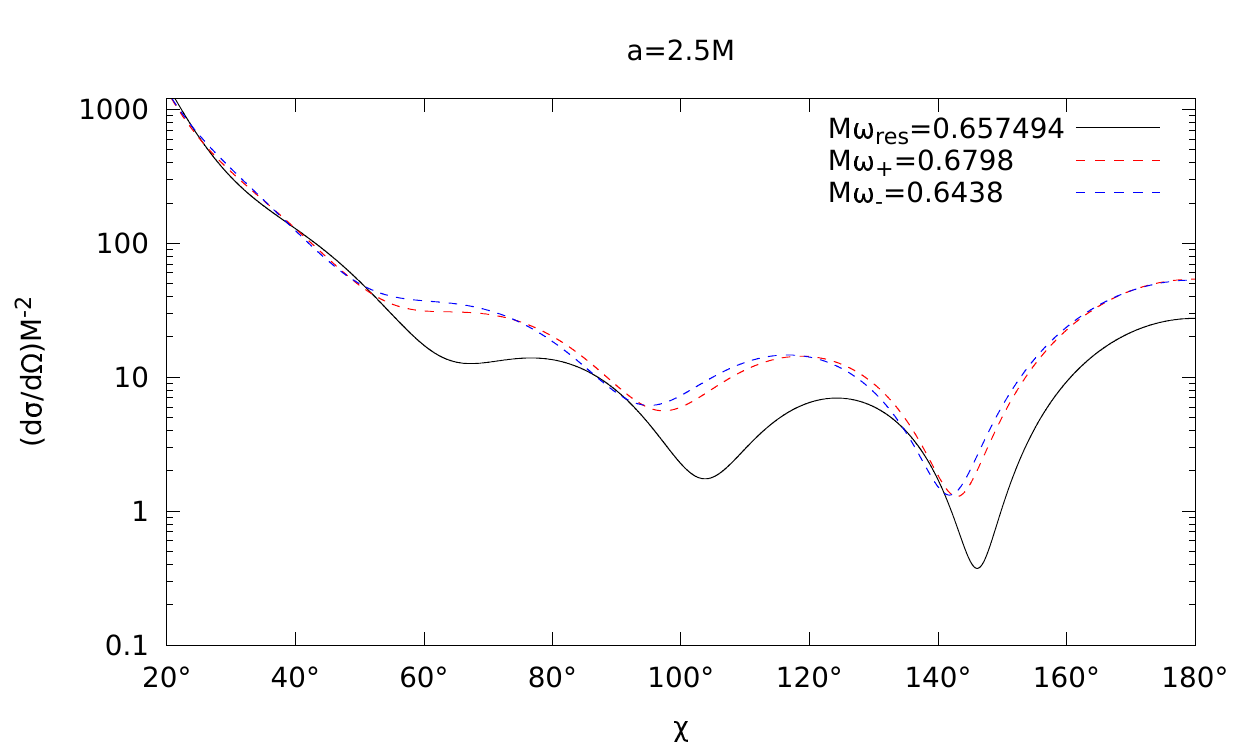}}
\caption{The scalar differential scattering cross section of traversable wormholes, described by the Simpson-Visser metric, for selected resonant frequencies $\omega_{\text{res}}$ (solid lines). We also show the differential scattering cross section for one frequency slightly larger ($\omega_+$) and one slightly smaller ($\omega_-$) than the resonant one.}
\label{WMHL-Ressonance}
\end{figure}

In Ref.~\cite{SV-1} we have shown that the absorption spectrum of the wormhole case presents resonant frequencies $\omega_{\text{res}}$, in which the absorption cross section exhibits sharp peaks~\cite{dmoc2019,msdc2018}. With this result in mind, we raise the following question: What happens to the scattering cross section at the resonant frequencies? 
In Fig.~\ref{WMHL-Ressonance} we show the scalar differential scattering cross section of traversable wormholes for some resonant frequencies $\omega_{\text{res}}$, computed in Ref~\cite{SV-1}, as well as for values of the frequency slightly larger ($\omega_+$) and slightly smaller ($\omega_-$) than the resonant frequencies. 
We note that the scalar differential scattering cross section substantially decreases at a resonant frequency, when compared to slightly different frequencies. 
The decreasing of the differential scattering cross section is more evident for large scattering angles.

\section{Final remarks}
\label{final remarks}
We have studied the scattering of scalar waves in the Simpson-Visser geometry that interpolates between the Schwarzschild black hole, regular black hole and traversable wormhole solutions. The general form of this geometry allows us to compare the scattering of black holes and wormholes by varying the interpolation parameter.

We analyzed the classical scattering cross section, which can be obtained by studying null geodesics in the Simpson-Visser geometry. We found an analytical expression for the classical scattering cross section in the limit of small scattering angles. Our analytical expression shows that the contribution from the interpolation parameter appears only in quadratic order on the classical scattering cross section. Thus, for small angles, the classical scattering cross section is predominantly given by the well-known Schwarzschild result. 

We have applied the semiclassical glory scattering formula to the black hole case. We obtained that the critical impact parameter of backscattered waves ($b_g$) slightly increases as we increase the interpolation parameter $a$. As a consequence, the interference fringes get slightly narrower as we increase $a$. We confirmed this behavior with the full numerical approach, which agrees very well with the glory approximation in the corresponding regime.

We applied the partial waves approach in order to compute numerically the differential scattering cross section of scalar waves in the Simpson-Visser geometry. We presented a selection of our numerical results for the black hole case, as well as for the wormhole case, and discussed the differences between the two situations.  

For the black hole case we found that the differential scattering cross section, for fixed $M\omega$ and $0\leq a<2M $, is very similar, although not equal, to the Schwarzschild results. The major difference arises for high angles, whereas for small angles the results are almost indistinguishable. The scattering cross section results for Simpson-Visser black holes reinforce the conclusions obtained in Refs.~\cite{SV-1,SV-5}, i.~e. that the Simpson-Visser black hole mimics the Schwarzschild solution in several aspects. We have shown that the full numerical results for the scattering cross section oscillates around the classical scattering cross section due to the interference of waves that orbit the black hole in opposite senses.

For the wormhole case we found that the differential scattering cross section presents a divergence in the forward direction ($\chi=0^\circ$) and also a glory in the backward direction ($\chi=180^\circ$), similarly to the black hole case. The partial waves numerical results oscillate around the classical scattering cross section. Moreover the differential scattering cross section for different values of $a$ and fixed $M\omega$ can be quite distinctive. Due to the presence of a potential well at the wormhole throat, the absorption cross section presents sharp peaks for some resonant frequencies $\omega_{\text{res}}$~\cite{SV-1,dmoc2019,msdc2018}. We investigated the behavior of the differential scattering cross section for frequencies equal to the resonant ones. We concluded that the differential scattering cross section is comparatively lower at the resonant frequencies. By comparing with results for frequencies slightly different from the resonant ones, we noticed that the relatively lower result is more evident for large scattering angles, while for small scattering angles they are essentially the same, since the forward divergence dominates in this regime.

\begin{acknowledgements}
The authors thank 
Funda\c{c}\~ao Amaz\^onia de Amparo a Estudos e Pesquisas (FAPESPA), 
Conselho Nacional de Desenvolvimento Cient\'ifico e Tecnol\'ogico (CNPq) and Coordena\c{c}\~ao de Aperfei\c{c}oamento de Pessoal de N\'{\i}vel Superior (Capes) - Finance Code 001, for partial financial support. We are grateful to Carlos Herdeiro and Pedro Cunha for useful discussions.
This research has also received funding from the European Union's Horizon 2020 research and innovation programme under the H2020-MSCA-RISE-2017 Grant No. FunFiCO-777740.
\end{acknowledgements}

{}
\end{document}